\documentclass[aps,prl,floatfix,twocolumn,groupedaddress,showpacs]{revtex4}

\usepackage{hyperref}
\usepackage{amsmath}
\usepackage{epsfig, graphicx}

\begin{document}
\newcommand{\simgt}{\,\hbox{\lower0.6ex\hbox{$\sim$}\llap{\raise0.6ex\hbox{$>$}}}\,}
\newcommand{\simlt}{\,\hbox{\lower0.6ex\hbox{$\sim$}\llap{\raise0.6ex\hbox{$<$}}}\,}
\newcommand{\aver}[1]{\langle {#1} \rangle}
\newcommand{\raw}[0]{\rightarrow}
\newcommand{\etaeff}[0]{\eta}
\newcommand{\bra}[1]{\ensuremath{\left\langle {#1} \right|}}
\newcommand{\ket}[1]{\ensuremath{\left|  #1 \right\rangle}}

\title{Cavity sideband cooling of a single trapped ion}

\author{David R. Leibrandt}
\email[]{dleibran@mit.edu}
\author{Jaroslaw Labaziewicz}
\author{Vladan Vuleti\'{c}}
\author{Isaac L. Chuang}
\affiliation{Department of Physics \& Center for Ultracold Atoms\\ Massachusetts Institute of Technology, 77 Massachusetts Avenue, Cambridge, MA, 02139, USA}

\date{\today}

\begin{abstract}
We report a demonstration and quantitative characterization of one-dimensional cavity cooling of a single trapped $^{88}$Sr$^+$ ion in the resolved sideband regime.  We measure the spectrum of cavity transitions, the rates of cavity heating and cooling, and the steady-state cooling limit.  The cavity cooling dynamics and cooling limit of 22.5(3) motional quanta, limited by the moderate coupling between the ion and the cavity, are consistent with a simple model [Phys.~Rev.~A \textbf{64}, 033405] without any free parameters, validating the rate equation model for cavity cooling.
\end{abstract}


\maketitle


Cooling atoms, molecules, and even macroscopic objects to the quantum mechanical ground state of their motion represents an important step towards comprehensive control of large or complex quantum mechanical systems \cite{diedrich89,hamann97,roos00b,schliesser08}.  Of particular interest are cooling methods that do not destroy the quantum mechanical coherences in other degrees of freedom.  For example, in trapped ion quantum information processing \cite{wineland98}, it is necessary to cool the motion of the ions to the ground state without affecting the internal state where the quantum information is stored.  This has been accomplished by sympathetic cooling of the processing ion using a laser cooled ion of another species \cite{home09}.

An all optical method that may allow one to cool a particle without causing decoherence of the internal states is cavity cooling.  When a particle is illuminated with monochromatic laser light, the spectrum of the scattered light is distributed around the incident frequency due to the particle's motion. If an optical resonator is used to enhance selected portions of the emission spectrum, such that the average emission frequency is larger than that of the incident light, energy conservation implies that the particle is cooled in the scattering process (cavity cooling) \cite{horak97,vuletic00,vuletic01}. Unlike conventional Doppler cooling \cite{haensch75}, cavity cooling does not require the incident light to be matched to an atomic resonance, and can in principle be performed with non-resonant light. As light scattered far off resonance carries no information about the atom's internal state \cite{cline94,ozeri05}, such a process can potentially cool an atom without destroying a quantum superposition of internal states.

Cavity cooling relies on the frequency discrimination provided by the resonator, and hence cooling to the quantum mechanical ground state of the external trapping potential is possible only when the trap frequency $\omega$ exceeds the cavity linewidth $\kappa$ (cavity resolved sideband regime) \cite{vuletic01}. In the resolved sideband regime cooling is achieved by tuning the resonator an amount $\omega$ to the blue of the incident light, such that scattering events into the cavity correspond to transitions $\ket{n} \raw \ket{n-1}$ that lower the motional quantum number $n$. The steady-state temperature is then set by the competition between cooling by photons scattered into the cavity and recoil heating by photons scattered into free space \cite{vuletic01}. Denoting the probability of scattering into a resonant cavity relative to free space by the cooperativity $\eta$, the steady-state average vibrational quantum number $n_{\infty}$ in the resolved sideband regime $\kappa \ll \omega$ is given by \cite{vuletic01}
\begin{equation}
n_{\infty} = \left( \frac{\kappa}{4 \omega} \right)^2 + \frac{C}{\eta} \left[1 + \left( \frac{\kappa}{4 \omega} \right)^2 \right],
\label{Eq:SteadyState}
\end{equation}
where $C$ is a dimensionless factor of order unity that depends on the cooling geometry. Thus in the strong-coupling regime $\eta \gg 1$ cooling to the motional ground state is possible, while for moderate coupling $\eta \lesssim 1$ the cooling is limited by the cooperativity to $n_{\infty}=C/\eta$.

Cavity cooling has been demonstrated in the weak confinement regime $\kappa \gg \omega$ with single trapped atoms without directly measuring the atomic temperature \cite{maunz04,nubmann05,fortier07} and with atomic ensembles in a different parameter regime of collective coupling \cite{chan03,black03}. In this work, using a single trapped $^{88}$Sr$^+$ ion in the resolved sideband regime, we measure for the first time cavity heating and cooling rates and the steady-state cooling limit, and observe parameter-free agreement with a rate equation model for cavity cooling \cite{vuletic01}.

\begin{figure}
\includegraphics[width=3.375in]{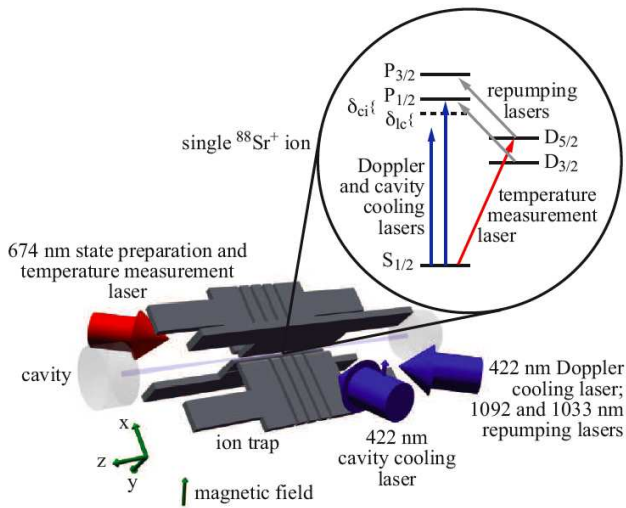}
\caption{\label{fig:ExperimentSchematic} (color online) Schematic of the experimental setup.  A single $^{88}$Sr$^+$ ion is confined in the center of a linear RF Paul trap and coupled to an optical resonator oriented along the trap axis.  The inset shows the ion energy levels (solid lines) and the cavity resonance (dashed line).}
\end{figure}

This experiment (Fig.~\ref{fig:ExperimentSchematic}) builds on the pioneering cavity cooling work with neutral atoms in the weak confinement regime \cite{chan03,black03,maunz04,nubmann05,fortier07} using the exquisite experimental control and strong confinement of trapped ions \cite{guthohrlein01,mundt02,russo08,herskind09}.  A single $^{88}$Sr$^+$ ion is confined in a linear RF Paul trap with motional frequencies $\omega_{x, y, z} = 2 \pi \times (1.45, 1.20, 0.87)$~MHz.  The cavity is 5~cm long and has a finesse $F = 2.56(1) \times 10^4$, resulting in a cavity linewidth (energy decay rate constant) $\kappa = 2 \pi \times 117(1)$~kHz.  The cavity is typically detuned by a few tens of MHz from the S$_{1/2} \leftrightarrow$ P$_{1/2}$ transition, which has a wavelength $\lambda = 422$~nm and a population decay rate $\Gamma = 2 \pi \times 20.2$~MHz.  The optical cavity is oriented along the ion trap axis and the 422 nm cavity cooling laser is perpendicular to the ion trap axis.  The additional 422 nm Doppler cooling laser, 1033 nm and 1092 nm repumpers, and 674 nm state preparation and temperature measurement laser are all at 45 degrees to the trap axis such that they have projections along all of the ion motional principal axes.  A 4.1 G magnetic field is applied to define the quantization axis.  The cooperativity at an antinode of the cavity for a two-level atom is $\eta_0 = 24 F / ( \pi k^2 w_0^2) = 0.26$, where $k=2\pi/\lambda$ is the wavenumber and $w_0 = 57.9(6)$~$\mu$m the radius of the cavity TEM$_{00}$ mode waist \cite{vuletic01}.

\begin{figure}
\includegraphics[width=3.375in]{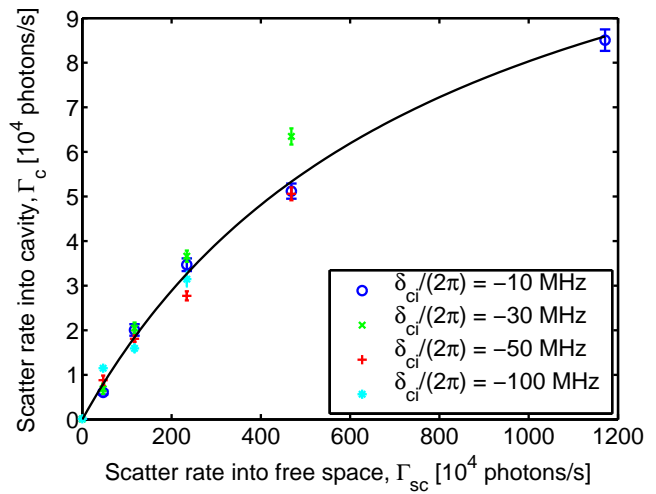}
\caption{\label{fig:Cooperativity} (color online) Photon scattering rate $\Gamma_c$ into the resonant cavity ($\delta_{lc} = 0$) as a function of the scattering rate into free space $\Gamma_{sc}$ for several values of cavity-ion detuning $\delta_{ci}$.  Each data point is measured by preparing the ion at fixed temperature by Doppler cooling for 200~$\mu$s, then measuring the photon scattering rates for 50~$\mu$s.  The line is a fit to the form $\Gamma_c = \etaeff \Gamma_{sc} / (1 + \Gamma_{sc} / \Gamma_{sat})$ with fit parameters $\etaeff = 0.018(4)$ and $\Gamma_{sat} = 8(2) \times 10^6$~photons/s.}
\end{figure}

The effective cooperativity is reduced from the cooperativity of a two-level atom $\eta_0$ by several effects, including the reduced dipole matrix element for $\pi$ polarized light on the $^{88}$Sr$^+$ S$_{1/2} \leftrightarrow$ P$_{1/2}$ transition (0.31), the error in positioning the ion in the center of the cavity mode (0.89), the non-zero laser linewidth (0.82), and the ion's thermal motion (0.32 for our typical temperature).  The combination of these independently measured effects results in a calculated effective cooperativity $\etaeff = 0.019$.  We determine $\etaeff$ experimentally by comparing the photon scattering rate by the ion into the resonant cavity $\Gamma_c$ (detuning between incident laser and cavity $\delta_{lc} = 0$) to the scattering rate $\Gamma_{sc}$ into free space.  The free space scatter is calibrated using a pulsed optical pumping experiment where the 422~nm and 1092~nm lasers are applied in succession and the mean number of 422~nm photons detected during the 1092~nm laser pulses gives the collection efficiency.  Fig.~\ref{fig:Cooperativity} shows the results for several values of the cavity-ion detuning $\delta_{ci}$ with the ion located at an antinode of the cavity standing wave.  The slope $d\Gamma_c/d\Gamma_{sc}$ at small values of $\Gamma_{sc}$ is the effective cooperativity $\etaeff = 0.018(4)$, consistent with our calculated value.  The saturation behavior of Fig.~\ref{fig:Cooperativity} at high free-space scattering rate is due to the finite repumping rate from the metastable D$_{3/2}$ state, which frequency-broadens the scattered light, and reduces the spectral overlap with the cavity mode.

\begin{figure}
\includegraphics[width=3.375in]{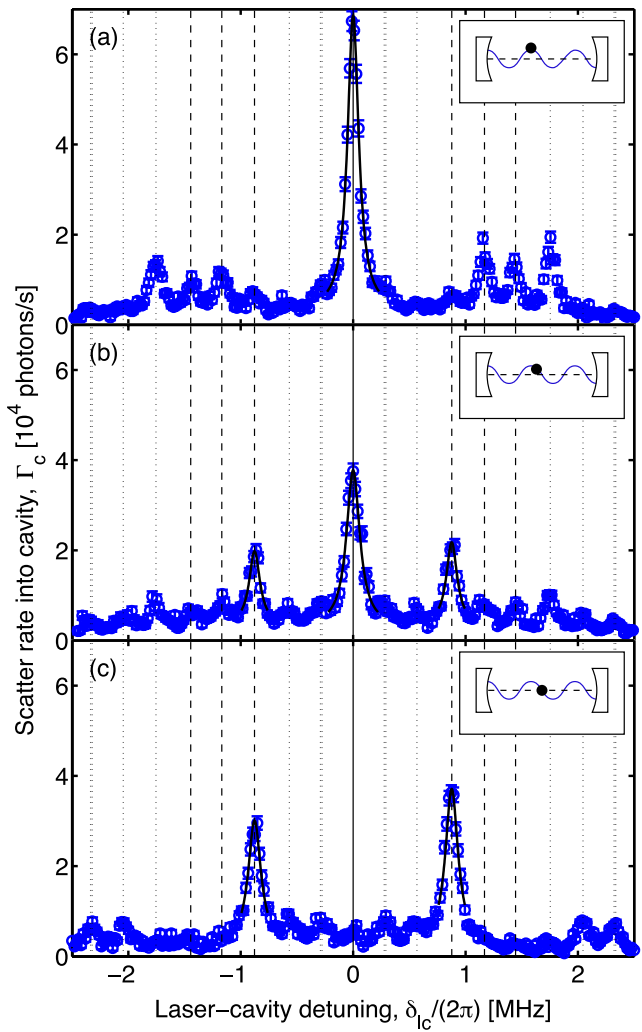}
\caption{\label{fig:Spectra} (color online) Photon scattering rate into the cavity as a function of laser-cavity detuning $\delta_{lc}$ with the ion located at a cavity standing-wave antinode (a), halfway between a node and an antinode (b), and at a node (c).  Each data point is measured by preparing the ion at fixed temperature by Doppler cooling for 200~$\mu$s, then measuring the rate of scattering photons from the cavity cooling beam into the cavity for 50~$\mu$s with $\delta_{ci}/(2 \pi) = -60$~MHz and $\Gamma_{sc} = 1.2 \times 10^7$~photons/s.  The solid, dashed, and dotted vertical lines are at the carrier, first-order motional sideband, and second-order motional sideband transition frequencies, respectively.  The curves are Lorentzian fits with linewidths consistent with the combined linewidth of the cavity and laser.}
\end{figure}

Due to the symmetries of the light-ion interaction, there are selection rules for the motional sideband transitions that depend on the location of the ion in the cavity standing wave \cite{russo08}.  The cavity scattering process is a two-photon transition where a photon is absorbed from the traveling-wave cooling laser and another photon emitted into the standing-wave mode supported by the cavity. The momentum  transfer in the scattering process enables the coupling between different ion motional states. Transitions which change the vibrational quantum number along the $x$ or $y$ directions perpendicular to the resonator ($\Delta n_{x,y} \neq 0$, enabled by the momentum of the absorbed photon) are strongest when the ion is located at a cavity standing-wave antinode, since this position corresponds to the strongest ion-cavity coupling. On the other hand, for vibration-changing transitions along $z$ that rely on the momentum of the emitted photon, the symmetry of the standing-wave field allows $\Delta n_z$-even processes ($\Delta n_z = 0,\pm 2, ...$) to occur only at an antinode, and $\Delta n_z$-odd processes ($\Delta n_z = \pm 1, \pm 3, ...$) only at a node. Consequently the relative strength of different motional transitions depends on the location of the ion in the cavity, as shown in Fig.~\ref{fig:Spectra}. When the laser-cavity detuning $\delta_{lc}$ is varied at fixed cavity-ion detuning $\delta_{ci}/(2 \pi) = -60$~MHz, the scattering into the cavity reveals all three first-order motional sideband transitions ($\Delta n = \pm 1$), as well as some of the second-order motional sideband transitions. The best cooling along the $z$-direction as investigated here is achieved via the $\Delta n_z = -1$ transition when the atom is located at a node of the cavity field.

\begin{figure}
\includegraphics[width=3.375in]{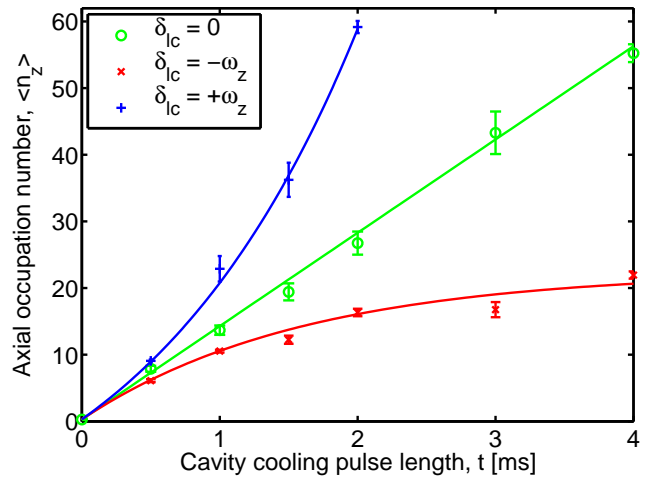}
\caption{\label{fig:CoolingDynamics} (color online) Cavity cooling dynamics.  The ion is sideband cooled to the three-dimensional motional ground state; a cavity cooling pulse with detuning $\delta_{lc} = 0$ (carrier), $\delta_{lc} = -\omega_z$ (red axial sideband), or $\delta_{lc} = +\omega_z$ (blue axial sideband) is applied; and the mean number of motional quanta in the $z$ mode is measured.  The three lines are a simultaneous fit to the model described in the Supplementary Information with fit parameters $n_0 = 0.30(6)$, $\Gamma_{sc} = 2.87(2) \times 10^6$ photons/s, and $\etaeff = 0.0148(2)$.  The reduced $\chi^2$ of the fit is 1.7.}
\end{figure}

We investigate one-dimensional cavity cooling along the $z$ direction by measuring separately the recoil heating rate, as well as the cavity cooling and heating rates for pumping on the cavity red ($\delta_{lc} = -\omega_z$) and blue ($\delta_{lc} = +\omega_z$) motional sidebands, respectively.  To realize a situation that allows simple quantitative comparison with the theoretical model for cavity cooling \cite{vuletic00,vuletic01}, we prepare the ion in its motional ground state by standard sideband cooling on the narrow S$_{1/2}, m = -1/2 \leftrightarrow$ D$_{5/2}, m = -5/2$ transition.  We then apply the cavity cooling laser for a variable time $t$ with detunings $\delta_{lc} = 0$ or $\delta_{lc}=\pm \omega_z$ and $\delta_{ci}/(2 \pi) = -10$~MHz.  Finally, the mean vibrational quantum number $\aver{n_z}$ is determined by measuring the Rabi frequencies of the red and blue motional sidebands of the S$_{1/2}, m = -1/2 \leftrightarrow$ D$_{5/2}, m = -5/2$ transition \cite{wineland98}.  This cavity-ion detuning is near the optimum value for conventional Doppler cooling, but the geometry of the setup dictates that the cavity cooling laser Doppler cools the $x$ and $y$ motional modes to maintain them at $\aver{n_{x,y} }\simlt 10$ but does not Doppler cool the $z$ motional mode.  The ion position is locked to a node of the cavity standing wave for this measurement by applying DC compensation voltages to the trap electrodes.  The recoil heating rate is the slope $d\aver{n_z}/dt$ for $\delta_{lc} = 0$ (Fig.~\ref{fig:CoolingDynamics} green line), and the cavity cooling and heating rates are the differences of the slopes $d\aver{n_z}/dt$ for $\delta_{lc} = \pm \omega_z$ (Fig.~\ref{fig:CoolingDynamics} red and blue lines) and for the recoil heating rate.  The signature of cavity cooling is that the temperature after pumping on the cavity red sideband is smaller than the temperature after pumping on the cavity carrier, which is smaller than the temperature after pumping on the cavity blue sideband. Cavity cooling ($\delta_{lc} = -\omega_z$) counteracts recoil heating by free-space scattering, and results in a finite steady-state vibrational quantum number $n_{\infty} \approx 20$.

We fit the cavity cooling dynamics to a rate equation model parameterized by the initial mean occupation number $n_0$, the free space scattering rate $\Gamma_{sc}$, and the effective cooperativity $\etaeff$ \cite{vuletic01} (see the Supplementary Information for a description of the model).  The data in Fig.~\ref{fig:CoolingDynamics} fits the model with fit parameters $n_0 = 0.30(6)$, $\Gamma_{sc} = 2.87(2) \times 10^6$~s$^{-1}$, and $\etaeff = 0.0148(2)$.  These values of the fit parameters are consistent with those derived from independent direct measurements, so that the rate equation model is consistent with our data without any free parameters. The steady-state mean occupation number due to cavity cooling is limited by the relatively small cooperativity of the cavity to $n_{\infty}=22.5(3)$, consistent with the value calculated from the model for our parameters.


In conclusion, we have demonstrated one-dimensional cavity cooling of a single $^{88}$Sr$^+$ ion in the resolved sideband regime. While our small effective cooperativity prevents us from cooling to the motional ground state, we have observed clearly resolved motional sidebands in the cavity emission spectrum, and we have measured the cavity cooling and heating rates.  Our results validate the rate equation model proposed by Vuleti\'{c} et al.~\cite{vuletic01}, which predicts that it is possible to cavity cool atoms or ions to the motional ground state without decohering the internal state. This would require a large detuning of the incident laser compared to the atomic fine structure, a criterion which is easier to meet with light ions such as Be$^+$ \cite{ozeri05}.

Resolved sideband cavity cooling might also be used to cool large molecular ions to the motional ground state \cite{kowalewski07,lev08}.  While some species of molecular ions have been previously cooled using sympathetic cooling \cite{molhave00,ryjkov06,ostendorf06}, large mass ratios between the atomic cooling ions and the molecular ions prevent efficient sympathetic cooling of the molecular ions at temperatures near the motional ground state.  Resolved sideband cavity cooling could enable exciting new studies of large molecular ions in the quantized motional regime.

\begin{acknowledgments}
We thank S.~Urabe for providing the linear RF Paul trap used in this work and J.~Simon, M.~Cetina, and A.~Grier for advice.  This work is supported in part by the Japan Science and Technology Agency, the NSF, and the NSF Center for Ultracold Atoms.
\end{acknowledgments}

\section{Supplementary information}

For cavity cooling in the weak coupling regime $\etaeff \ll 1$, coherences \cite{zippilli05a,zippilli05b} decay rapidly, and rate equations are sufficient to describe the cooling \cite{vuletic01}.  For one-dimensional cooling along the cavity axis $z$, the rate of transitions from motional state $\ket{n}$ to $\ket{n - 1}$ is
\begin{equation}
 \frac{\Gamma_{sc} \etaeff \eta_{LD}^2 n}{1 + 4 \left( \delta_{lc} + \omega \right)^2/\kappa^2} \equiv R^- n,
\end{equation}
and the rate of transitions from motional state $\ket{n}$ to $\ket{n + 1}$ is
\begin{equation}
\begin{array}{r}
\frac{\Gamma_{sc} \etaeff \eta_{LD}^2 (n + 1)}{1 + 4 \left( \delta_{lc} - \omega \right)^2/\kappa^2} + \Gamma_{sc} C \eta_{LD}^2 + \dot{n}_{ext} \equiv R^+ (n + 1) + N^+,
\end{array}
\end{equation}
where $\Gamma_{sc}$ is the photon scattering rate into free space, the number $C$ is defined such that $C \eta_{LD}^2 \hbar \omega$ is the average recoil heating along the $z$ direction per free space scattering event, and $\dot{n}_{ext}$ is the heating rate along the $z$ direction due to environmental electric field fluctuations in quanta per second.  Here, $\eta_{LD}^2 = E_{rec}/(\hbar \omega)$ is the Lamb-Dicke parameter, as determined by the ratio of recoil energy $E_{rec}$ and trap vibration frequency $\omega$.  Note that these transition rates are only valid in the Lamb-Dicke regime $\eta_{LD}^2 \left< n \right> \ll 1$, which limits the applicability of this model to $\left< n \right> \ll 70$ for our experimental parameters.  The expectation value of the mean vibrational quantum number $\aver{n}_t$ evolves according to
\begin{equation}
\left< n \right>_t = n_0 e^{-W t} + n_{\infty} \left( 1 - e^{-W t} \right)
\end{equation}
for $\delta_{lc} = -\omega$ (cooling),
\begin{equation}
\left< n \right>_t = n_0 + \left( R^+ + N^+ \right) t
\end{equation}
for $\delta_{lc} = 0$, and
\begin{equation}
\left< n \right>_t = \left( n_0 + n_{\infty} + 1 \right) e^{W t} - \left( n_{\infty} + 1 \right)
\end{equation}
for $\delta_{lc} = +\omega$ (heating), where $n_0 \equiv \aver{n}_{t=0}$ and $n_{\infty} \equiv \aver{n}_{t \raw \infty}$ are the initial and steady-state value of $\aver{n}_t$, respectively.  The cavity cooling rate constant $W$ is given by
\begin{equation}
W = \frac{\Gamma_{sc} \etaeff \eta_{LD}^2}{1 + \kappa^2/ \left( 2 \omega \right)^2},
\end{equation}
and the steady state average occupation number $n_{\infty}$ is given by
\begin{equation}
n_{\infty} = \left( \frac{\kappa }{4 \omega} \right)^2 + \left[ \frac{C}{\eta} + \frac{\dot{n}_{ext}}{\Gamma_{sc} \etaeff \eta_{LD}^2 } \right] \left[ 1 + \left( \frac{\kappa }{4 \omega} \right)^2 \right] \ .
\end{equation}
For cavity cooling of the $z$ motional mode in our experiment, we calculate $C = 1/3$ (photons are scattered isotropically for a $J = 1/2 \leftrightarrow J' = 1/2$ transition), and measure independently $\dot{n}_{ext} = 17(2)$~s$^{-1}$. Thus, for our experimental parameters, the heating due to environmental field fluctuations is negligible ($\dot{n}_{ext} \ll \Gamma_{sc} \etaeff \eta_{LD}^2$) and the expression for the steady-state occupation number reduces to Eq.~(1).


\end{document}